\documentclass{iau-JDSS}
\usepackage{graphicx}

\title[Astronomical Data Management] 
{Astronomical Data Management}

\author[Norris et al.]   
{ Ray Norris$^1$, Heinz Andernach$^2$, Guenther Eichhorn$^3$, \break
Fran\c{c}oise Genova$^4$, Elizabeth Griffin$^5$, Robert Hanisch$^6$, \break
Ajit Kembhavi$^7$, Robert Kennicutt$^8$, Anita Richards$^9$. }

\affiliation{$^1$CSIRO ATNF, PO Box 76, Epping, NSW 1710, Australia. 
email: Ray.Norris@csiro.au\\[\affilskip]
$^2$ Depto.\ de Astronom\'\i a, Univ.\ Guanajuato, Mexico.
email: heinz@astro.ugto.mx \\
$^3$ Astrophysics Data System, Smithsonian Astrophysical Observatory, 60 Garden St., MS-67, Cambridge, MA 02138, USA.
email: gei@cfa.harvard.edu\\
$^4$ CDS, UMR CNRS/ULP 7550, Observatoire Astronomique, 11 rue de l'Universit\'e, 67000 Strasbourg, France. email: genova@astro.u-strasbg.fr\\
$^5$ NRC Herzberg Institute of Astrophysics, 5071 West Saanich Road,
Victoria, British Columbia, Canada. V9E~2E7
email: Elizabeth.Griffin@hia-iha.nrc-cnrc.gc.ca\\
$^6$ Space Telescope Science Institute, 3700 San Martin Drive, Baltimore, MD 21218 USA. \break
email: hanisch@stsci.edu\\
$^7$ Inter-University Centre for Astronomy and Astrophysics, Pune, India.
\break email: akk@iucaa.ernet.in\\
$^8$ Institute of Astronomy, 
University of Cambridge, Madingley Road,
Cambridge. CB3 0HA, UK.
email: robk@ast.cam.ac.uk\\
$^9$ MERLIN/VLBI National Facility, University of Manchester, 
Jodrell Bank Observatory, Macclesfield, Cheshire SK11 9DL, U.K.
email: amsr@jb.man.ac.uk\\
}

\pubyear{2006}
\volume{Volume 14}  
\pagerange{1--10}
\date{?? and in revised form ??}
\jname{Highlights of Astronomy, Volume 14}
\editors{K.A. van der Hucht, ed.}
\begin{document}

\maketitle

\begin{abstract}
We present a summary of the major contributions to the Special Session on Data Management held at the IAU General Assembly in Prague in 2006. While recent years have seen enormous improvements in access to astronomical data, and the Virtual Observatory aims to provide astronomers with seamless access to on-line resources, more attention needs to be paid to ensuring the quality and completeness of those resources.  For example, data produced by telescopes are not always made available to the astronomical community, and new instruments are sometimes designed and built with insufficient planning for data management, while older but valuable legacy data often remain undigitised. Data and results published in journals do not always appear in the data centres, and astronomers in developing countries sometimes have inadequate access to on-line resources. To address these issues, an ``Astronomers' Data Manifesto'' has been formulated with the aim of initiating a discussion that will lead to the development of a ``code of best practice'' in astronomical data management.

\end{abstract}

\firstsection 

\section{Introduction}

The last few years have seen a revolution in the way astronomers use data. Data centres such as ADS (Astrophysics Data System, http://adswww.harvard.edu/), CDS (Centre de Donn\'ees astronomiques de Strasbourg, http://cdsweb.u-strasbg.fr/), and NED (NASA/IPAC Extragalactic Database, http://nedwww.ipac.caltech.edu/) have transformed the way we access the literature and carry out our research. Archival research with space observatory data is becoming the dominant driver of new research and publications, while major ground-based mission archives are not far behind. Access to electronic data and publications has brought front-line research capabilities to all corners of the developed world, and a growing number of archives from major telescopes are being placed in the public domain. Other success stories include the vigorous international development of the Virtual Observatory (VO), the revolutionary public data releases from individual astronomical projects, and the rapid dissemination of results made possible by forward-thinking journals, the ADS, and the astro-ph preprint server. All of these position astronomy as a role model to other sciences for how technology can be used to accelerate the quality and effectiveness of science. 
  
On the other hand, our management of astronomical data is still inadequate, to the detriment of our science. For example, there exist international pressures to surround our open-access databases in a morass of legal red tape, and we are poorly prepared to resist them (Norris, 2005). There remains a bottleneck between journals and data centres, so that a significant fraction of important data published in the major journals never appears in the data centres, and there is little provision for the preservation of the digital data underlying the results published in peer-reviewed journals. New instruments are still being built with little planning or budgeting for data management, so that while the instrument may technically perform well, the quality of the delivered science fails to meet expectations or capacity. While astronomers in the developed world revel in instant access to data and journals, their developing-world colleagues still rely on photocopied preprints. Valuable legacy data which might prove crucial to the understanding of the next supernova lie undigitised and inaccessible in some remote storage room, at the mercy of natural hazards and human ignorance. 

In many astronomical institutions, data management as a discipline is not yet taken seriously. For example, an astronomer making a large database publicly available is not given the recognition that is given to the author of a paper, even though the database may effectively attract far more "citations" than the paper. 

In an effort to raise awareness of these issues, and to work towards a policy of ``best-practice'' astronomical data management, a Special Session on ``Data Management'' was held at the IAU General Assembly in Prague in 2006. In addition, a lively and productive electronic discussion (http://www.ivoa.net/twiki/bin/view/Astrodata ) took place over several months preceding the IAU General Assembly. This paper presents a summary of the contributions to this Session, including not only the oral and some poster presentations, but also key points from the e-discussion. It is co-authored by the main contributors to the oral sessions and to the preceding e-discussion, and represents a consensus view. Inevitably, it falls short of conveying the spirited discussion that livened the meetings.

\section{The Virtual Observatory}

\subsection{Overview of the Virtual Observatory}

The VO aims to provide astronomers with seamless access to on-line resources. A good overview of its present status is provided by the proceedings of Special Session 3 in these {\it Highlights of Astronomy}. Although there are many national VO projects, each working in a specific context with its own goals, strongly dependent on the local funding agencies, the VO is a world-wide, global endeavour, and all projects work together in the International Virtual Observatory Alliance (IVOA), which was formed in 2001. Among its main tasks, the IVOA defines the VO interoperability standards, including the standards for resource registries, query language, data access layer, content description (semantics), and data models. Most of the essential VO standards are now ready or nearly ready, and the VO is now in transition from R\&D infrastructure definition to implementation by data centres. Many participants in the groups in charge of defining the standards, formed by staff from national projects, are knowledgeable about data archives and on-line services, and care about providing useful and usable standards, with the aim that data and services can be published in the VO through a thin interface layer.

Because the VO has successfully developed these standards, together with a growing range of tools and services, it has attracted a huge visibility, interest, and respect in the IT (information technology) community. And yet perceptions in the broader astronomical community are mixed, ranging from those who enthusiastically use VO tools to generate science, to those who consider that its value and relevance have yet to be demonstrated. VO proponents acknowledge that some critical challenges remain (e.g. long-term curation, quality control, certification, intellectual property guidelines, version control) and a number of key tasks remain unfinished, but the momentum of VO development is steadily meeting this challenge.

\subsection{Data Centres in the VO}

The definition of a data centre can range from ``a place distributing observational data'' to a service such as the CDS or NED which also distributes information, tools, and value-added services. In the VO context, new types of data and service providers emerge, and it is more appropriate to define a ``data centre'' in terms of attributes such as {\it service to the community}, {\it added-value linked to expertise}, {\it sustainability}, and {\it quality}.

Many teams are willing to provide VO-compliant data and services in their domains of expertise. Key participants in the VO  include the ``classical'' data centres such as observatory archives, discipline-specific data centres, and data centres like CDS and NED which provide reference services and tools. In addition, a growing number of scientific teams are willing to participate by providing specific value-added services and tools in their domains of expertise. For instance, when the French VO performed a census in 2004, more than 40 teams planned to participate in VO-related actions, and most of these confirmed their activity in a census update two years later, indicating real commitment.

The community of VO data and service providers is therefore diverse both in the size of the teams, and in the context in which they work, and range from  large national or international agencies to small teams working in scientific laboratories. Many types of services are being implemented, such as: 
\begin{itemize}
\item observation archives, with a strong emphasis on ``science ready'' data,
\item value-added services and tools, with compilations, including additional data required for data interpretation, such as data on atomic and molecular lines,
\item theoretical services, with on-demand  services, or sets of pre-cooked modelling results,
\item software suites, in particular for data analysis,
\item specific services, to help solving specific science questions, and
\item full research environments.
\end{itemize}

One key objective for the VO projects in the coming years is to create a community of VO service providers, who will help data centres to use the VO framework, and gather their feedback from implementation. This is a very important role for the national VO projects, and IVOA has to take into account the implementation feedback. The functionality of the VO in this increasingly operational phase is illustrated by the schema of the Euro-VO, which has three facets interacting together: a Data Centre Alliance, a distributed Technology Centre in charge of infrastructure definition, and a Facility Centre (ESA, ESO and national projects) which provides general information and supports users.

\subsection{The future of the VO}

Since the advent of the Internet, astronomy has been at the forefront for provision and networking of on-line data and services. This has already produced a revolution in the way astronomers work, even if they do not always realize it and simply use the tools. The VO is the next step, providing new resources and seamless access to them. New data and tools are already here and will be continue to be added.

The VO development provides a strong incentive to observatories and scientific teams to make their data and services available to the whole community, so that many teams want to become VO data centres. This is excellent news, since it will increase the sharing of information and knowledge among the community. Data centres have certain requirements, including
\begin{itemize}

\item a {\it critical mass} adapted to the aims, 
\item {\it medium term sustainability}, which requires strong support from the funding agencies, and
\item {\it national/international scientific niche} to gain community support. 
\end{itemize}

Data centres also have significant responsibilities, including {\it curating data}, which comes with a large overhead for selecting, homogenizing, describing, and distributing data, and data centres must be sufficiently funded to perform these tasks. One has to keep in mind that data centres can be terminated and that it is critical for them to define a long term strategy, and to adjust it to the scientific evolution of astronomy, to technical evolution, and to the evolution of context, such as the development of the VO.

Why should data providers join the VO? They will have to care more about data quality and metadata, which means more work, but they will improve their service, and will have more occasions to collaborate with colleagues and build synergies between their services, and their visibility and usage statistics will increase significantly. The most difficult task will probably be to provide and maintain the service and to ensure quality, not to implement the VO framework!

Success of an operational VO network will ultimately be measured by customer participation and satisfaction, where the "customers" include both users and data providers. And while many VO elements, such as format standardization and user tools, are already in place, others are still being addressed. These remaining challenges include long-term data access, data quality and curator certification, version control, histories, and intellectual property standards. These require a shift in focus away from technological tools towards a suite of data management processes.

\section{Open Access and observatory archives}
\subsection{Open Access}

Because the advance of astronomy frequently depends on the comparison and merging of disparate data, it is important that astronomers have access to all available data on the objects or phenomena that they are studying. Astronomical data have therefore always enjoyed a tradition of open access, best exemplified by the astronomical data centres, which provide access to data for all astronomers at no charge. There exist a number of exceptions to this open access tradition, some of which are widely supported, such as the initial protection of observers' data by major facilities. 

At the 2003 IAU General Assembly a resolution was adopted which says, broadly, that publicly-funded archive data should be made available to all astronomers. This is aligned with ICSU (International Council for Science) and OECD (Organisation for Economic Cooperation and Development) recommendations, and may be regarded as a first step towards articulating the principles by which the astronomical community would like to see its data managed. Since then, a number of observatories, notably the European Southern Observatory (ESO), have embraced an open-access policy, but there remain a number of observatories that have not yet made their archival data publicly available, typically because of resource constraints. There also remain a few observatories (primarily privately-funded) which allow data archive access only to affiliated scientists, while still benefiting from the open access policies of other institutions. 

The adoption of an open-access policy is not just for the public good. Roughly three times as many papers (and citations) result from data retrieved from the Hubble archive as those based on the original data (Beckwith, 2004). In the parallel case of IUE (International Ultraviolet Explorer) spectra, five times as many publications resulted from archive data (Wamsteker \& Griffin, 1995). So, in principle, observatories might quadruple their science by making their archive data public. Since the funding for most major observatories depends on performance indicators such as publications and citations, it may be an expensive decision for an observatory not to adopt an open-access policy.

\subsection{Data Needs for New Telescopes}

Part of the success of modern astronomy can be attributed to astronomers who continue to strive for bigger and better instruments. But as  plans are developed for a new telescope, data processing and management are sometimes neglected. However, half the cost of a modern ground-based telescope is typically in the software and data processing. These need to be planned and developed at the same time as the hardware, rather than leaving it to graduate students or support staff to figure out when the data arrive. This may seem obvious, especially to those major projects that already routinely follow this practice. However, some projects have not shown such foresight, resulting in instruments which perform well technically, but which have not delivered the expected scientific impact. To avoid this, it is vital to think about these issues before, rather than after, the telescope is funded and built. 

\section{Journals and Data}
\subsection{The changing face of astronomical publications}

arXive/astro-ph is now the primary channel for disseminating new research results, and ADS is now the primary channel for accessing published papers. Electronic editions have become the main journals of record, and the days of paper journals are numbered. The primary journals (A\&A, AJ, ApJ, MNRAS) are adapting to this new publishing paradigm, but the future of commercial and small journals is unclear. Meanwhile, astronomical monographs and conference proceedings generally remain locked in the old paradigm, and consequently their impact is declining. Other components of "grey literature", the observatory reports and technical papers, are locked out of the new paradigm, and are being lost.

A further consequence of this changing paradigm is that the current business model for astronomical publications is being challenged. Most astronomers accept the need for high-quality peer-reviewed journals, while searching for ways in which they can be improved, and made cheaper. But there is a growing demand for open access, or free, journals, although it has yet to be demonstrated how an open-access journal can afford to maintain the quality that we have come to expect from our mainstream journals. Thus, the future business model for peer-reviewed publication is unclear.

In addition to accessing the journal article itself, astronomers are demanding better links between publications and data, where the term ``data'' is taken to include primary observational data, published results based on those data, and graphical representations of results. Astronomers' requirements vary from field to field, and include:

\begin{itemize}
\item Tables published in a journal should be accessible by catalogue browsers such as VizieR.
\item Results published in the journal should appear in object-searchable or position-searchable databases such as NED or SIMBAD.
\item Readers should be able to click on an object in a journal to obtain more information about that object from a database such as NED or SIMBAD
\item Users of NED or SIMBAD should be able to trace a link back to a refereed publication which validates and authenticates the data.
\item Links should be given in a publication to an archive containing its source data.
\end{itemize}

Both the journals and the data centres are actively addressing these issues. For example, a collaboration between the American Astronomical Society (AAS) and the Astronomical Data Center Executive Committee (ADEC) has put in place a system that allows authors to specify data that they used in their articles. This information is then processed by the publisher and used to link from articles to on-line data, both in the journals, and the ADS. It is hoped that other publishers and data centres world-wide will participate in this system to provide such links world-wide.

Although the journals are embracing opportunities engendered by the new technologies such as active links from electronic journals to the data centres, the metadata (and for that matter, the quality of error-checking in the tables themselves) provided by authors are not currently sufficient to enable completely automated transfers of the results from journal to data centre. Consequently, data centres often have to go through published tables by hand. Their capacity to do so is strongly limited by available resources, and so a significant fraction of published results never appears in a data centre, or does so only after a period of some years.

For example, Andernach (2006) has conducted a case study of over 2000 published articles, for which he collected or restored (via OCR) the electronic tables they contain, and finds that typically only about 50\% of results published in journals ever appear in the data centres, and lists some surprising and significant omissions. Strangely, this fraction did not appear to change significantly as journals changed from print-only to electronic formats. 

One solution to this data bottleneck would presumably be to increase funding for the data centres to enable them to employ more staff to transcribe and interpret the journal data, but the finite resources available make this option unlikely. An alternative option is to define formats and metadata that are author-friendly, journal-friendly, and data centre-friendly, and define the data sufficiently well. Then, if an author chooses to supply these metadata, and certifies that he has checked the data using appropriate tools (many of which are already available), they could be imported automatically into the data centres. This effectively redistributes the transcription workload from the data centres to the authors, and necessarily entails more work for authors. However, the authors will benefit from the greater scientific impact and the higher citation rate that will result from their data being in the data centres. In many cases, the paper itself will benefit from this further level of checking, which will remove the errors that are still too common in published papers, which therefore require checking by data centre staff before their data are accepted by the data centres.

There is no widespread agreement whether such a system can ever be made to work reliably without reducing the quality of the data in the centres. Some astronomers are concerned that using such tools would allow more errors to remain undetected when data are deposited in data centres. Others argue that this concern is outweighed by the advantages of easier access to data. Data centres experienced in handling author-provided data and metadata have expressed doubt that quality can be insured without a final stage of checks by data centre or journal experts. A pilot study seems justified to explore the feasibility and examine whether results based on data published in this way are in fact less reliable. 

\subsection{Digital Data Preservation for Astronomy Journals}

Astronomers are producing and analysing data at ever more prodigious rates. NASA's Great Observatories, ground-based national observatories, and major survey projects have archive and data distribution systems in place to manage their standard data products, and these are now interlinked through the protocols and metadata standards agreed upon in the Virtual Observatory. However, the digital data associated with peer-reviewed publications are only rarely archived. Most often, astronomers publish graphical representations of their data but not the data themselves. Other astronomers cannot readily inspect the data to either confirm the interpretation presented in a paper or extend the analysis. Highly processed data sets reside on departmental servers and the personal computers of astronomers, and may or may not be available a few years hence. 

A project led by Hanisch is investigating ways to preserve and curate the digital data associated with peer-reviewed journals in astronomy. The technology and standards of the VO provide one component of the necessary technology. A variety of underlying systems can be used to physically host a data repository, and indeed this repository need not be centralized. The repository, however, must be managed and data must be documented through high quality, curated metadata. Multiple access portals must be available: the original journal, the host data centre, the Virtual Observatory, or any number of topic-oriented data services utilizing VO-standard access mechanisms. 

The near-term goal of this project is to implement an end-to-end prototype digital data preservation facility using astronomy scholarly publications as a test-bed. Astronomy is an ideal discipline to start with, as most data are available in a single standard format (FITS), the community is small and highly aware of e-publishing, and there are few restrictions on data access and exchange. The prototype will be implemented using commodity open-source technologies and will utilise the infrastructure already being developed by the VO in order to minimize development costs and maximize flexibility. Specific development tasks include metadata definition, evaluation and selection of a content management tool (Fedora, DSpace, etc.), deployment of storage applications and layered storage management software (VOSpace), and adapting the publication process for data capture. By implementing a prototype, it is hoped to understand operational costs and thus be able to develop a long-term business plan for the preservation of peer-reviewed journal content and the associated supporting data. The availability of such facilities for digital data preservation will undoubtedly lead to changes in policies affecting data access. Peer pressure may initially encourage researchers to contribute their data to the repository, but eventually such contributions might become mandatory. It seems likely that published papers having digital data available will be more heavily used, and thus more heavily cited, than papers lacking such data. In the longer term it will be important to evaluate the impact on scientific productivity through citation analyses and community feedback. 

\section{Challenging the Digital Divide} 

The ``Digital Divide'' refers to the widening gulf between those who have high-bandwidth access to information, data, and web services, and those who do not. For those who do not, their lack of access results in even more disadvantages, making it even less likely that they will gain access in the future. 

Astronomers in developing countries are better positioned than their colleagues in some other disciplines, because the astronomical data centres maintain immense databases, and electronic archives of scientific periodicals, while the latest research is available through preprints on astro-ph. Furthermore, some of the leading astronomical journals provide free or cheap access to astronomers in developing countries. In the future, the situation is set to improve further, as powerful Virtual Observatory tools will provide even better access to astronomical data. Meanwhile, facilities such as SALT and GMRT are already demonstrating the feasibility of building leading-edge facilities, complete with well-managed data archives, in developing countries. However, many astronomers in these countries lack the bandwidth, expertise and the environment to make use of these riches. Further obstacles include resistance to the use of new concepts and tools, and reservations about exposing hard-won data to international access.

For example, India possesses several research institutes with state-of-the-art facilities, including access to high bandwidth, databases, literature and computing facilities. The Indian software industry is one of the most successful in the world, and yet very few Indian astronomers make extensive use of archival data for large scientific projects, and little attention is given to software aspects of large astronomical projects. As a result, the Information Technology prowess of India in the business domain has not been exploited by the scientific community, astronomical data from Indian observatories have not been archived and made available to the community, and India remains on the wrong side of the digital divide.

The situation in Africa, which does not have the technological advantages available to India, is even worse. Most institutions do not have good internet bandwidth, and ADS access statistics show that although African ADS usage is increasing, African astronomers are not yet taking full advantage of the available digital information. 

However, because of a number of initiatives, the situation seems to have improved slightly in the past few years. For example, associates and their students from all over India are funded to spend a few months every year at the Inter-University Centre for Astronomy and Astrophysics, where they develop their own research programs and set up collaborations. The resulting technology and expertise are transferred to the universities, helped by the decreasing cost of personal computers.
 
This shows that such efforts are producing results and need to be continued and supported as much as possible. In particular, the Indian experience could be replicated elsewhere in the developing world. Although the digital divide problem extends over all disciplines, astronomy is well-positioned to lead the charge to challenge this divide. Astronomers in the developing world could help build archives, develop software and provide much needed human resources, using a platform provided by the Third World Astronomy Network.

\section{Safeguarding Data}

There are many reasons, both scientific and economic, why a properly-managed data archive is an essential facility in astronomy. Most modern astronomers agree with the principle of archiving data for the wider community benefit, but in practice our achievements are patchy, particularly in the case of the preservation and accessibility of historic data. While the VO is currently focussing on modern space- and ground-based data that were recorded digitally, much less attention is being paid to astronomy's rich legacy of photographic observations, some of which date back to the late 19th century. 

The value of such data to modern science has been demonstrated repeatedly, through studies of very long-period variability (something that is predicted, but scarcely figures even today in the astrophysics landscape), identifying and measuring non-recurrent events such as spectrum changes in AGB stars, refining small-body orbits (including those crucial near-earth objects), studying the pre-outburst phases of a supernova (such as was very fortunately possible for SN1987A), resolving important ambiguities or anomalies through more precise re-measurements of historic data, or in inter-disciplinary science such as measuring the Earth's ozone concentrations as extracted from historic stellar spectra. There are also many data sets on magnetic tape that were abandoned as incompatible technology moved ahead without them. Although the physical longevity of photographic data far outmatches that of tapes, resources for safeguarding them are necessarily in competition with those required to generate new data. It is therefore important to determine, as far as we can, what value to place on the historical archives, and to determine a workable solution for their long-term storage and digitization before we lose the opportunity to make that decision.

The migration of present-day digital data is now well orchestrated in data centres, so that as technology moves on, data are migrated seamlessly to new media or new formats. Outside the data centres, however, the problem remains. Astronomers and small observatories keep magnetic tapes, including DAT and Exabytes tapes, well beyond their recommended life. Few now have the technology to read a round magnetic tape or a 5 $\frac{1}{4}$-inch floppy. How long before an Exabyte, or even a CDROM or a DVD, becomes unreadable?
 
Astronomy needs to take a broader view of safe-guarding its data. The cost of recovering historic observations as a common-user resource is small compared to new installations or space missions. 

\section{The Astronomers' Data Manifesto}

In an attempt to raise awareness of these issues, and define the goal, the IAU Working Group for Astronomical Data proposed the following manifesto. This is intended not as a rigid declaration, but as a stimulus for discussion. It is hoped that such discussion will converge to a consensus on how the astronomical community would like to see its data managed.

``We, the global community of astronomy, aspire to the following guidelines for managing astronomical data, believing that they would maximise the rate and cost-effectiveness of scientific discovery. We do not underestimate the challenge, but believe that these goals are achievable if astronomers, observatories, journals, data centres, and the Virtual Observatory Alliance work together to overcome the hurdles.

\begin{enumerate}

\item  All significant tables, images, and spectra published in journals should appear in astronomical data centres. 

\item  All data obtained with publicly-funded observatories should, after appropriate proprietary periods, be placed in the public domain. 

\item  In any new major astronomical construction project, the data processing, storage, migration, and management requirements should be built in at an early stage of the project plan, and costed along with other parts of the project. 

\item  Astronomers in all countries should have the same access to astronomical data and information. 

\item  Legacy astronomical data can be valuable, and high-priority legacy data should be preserved and stored in digital form in the data centres. 

\item  The IAU should work with other international organisations to achieve our common goals and learn from our colleagues in other fields. ''

\end{enumerate}

\section{Conclusion}

The revolution in the way that astronomy manages its data has already resulted in enormous scientific advances. The potential for further advances in the future is even greater, but we need to have a clear vision and a clear goal if we are to succeed in realising that potential. In particular, the astronomical community needs to have a clear picture of what represents ``best practice'' in astronomical data management. Astronomy does not have any strategic data framework to provide policies or guidelines for astronomical data management, and is not therefore able to represent the interests of astronomical data management to external parties. For example, data quality, long-term accessibility, and provenance carry real costs but are critical requirements for success. 

The Virtual Observatory is a powerful tool that will enable us to make even more effective use of our data, but we should not regard it as a cure-all for our current deficiencies in data management. Whilst the VO is attempting to make major databases accessible to all astronomers, it cannot do so unless those databases are properly constructed and managed. Now that the infrastructure is in place we need to focus on building a user base and bringing in all key archives and collections.

These advances also have the potential to overcome the Digital Divide, but only if further initiatives enable open access to these facilities by astronomers in developing countries. Such initiatives are likely to be cost-effective, as the VO, electronic publication, and effective archives will enable science to tap an enormous intellectual base, with fresh ideas and approaches, which will benefit all of us.

\end{document}